\documentclass[]{cupconf}

\usepackage{graphicx}

\title[The stellar population of bulges ]
      {The stellar population of bulges}

\author[P.Jablonka]{\\ P. Jablonka} 

\affiliation{Ecole Polytechnique F\'ed\'erale de Lausanne \& Universit\'e
  de  Gen\`eve,\\ Observatoire, chemin   des Maillettes   51,  CH-1290
  Sauverny, Switzerland }

\begin{document}

\maketitle

\begin{abstract}

This review   summarizes the properties  of the  stellar population in
bulges as observed  in nearby or distant spiral  galaxies. It gives  a
particular emphasis to the comparison  with elliptical galaxies, when
possible.  The    criteria of sample  selection   and  choices in data
analysis are addressed when they may be involved in discrepant results
reached by different studies.

\end{abstract}

\firstsection

\section{Introduction}

\begin{figure}
\begin{center}
\includegraphics[width=2.5in,height=3.5in,angle=-90]{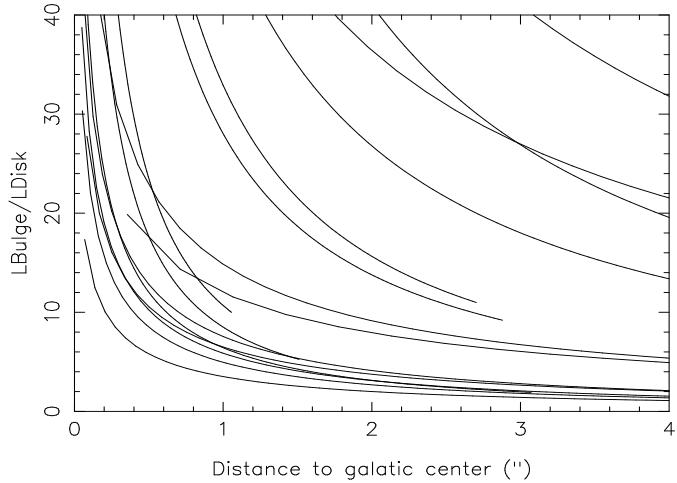}
\hfill
\caption{The ratio between the bulge and disk luminosities as a function of
  the distance from  the galaxy center for  a sample of nearby face-on
  galaxies.}
\label{LBLD}
\end{center}
\end{figure}

Studying bulges of  spirals is  not restrictive to a  particular
class of galaxies,  or even further to  their central  regions.  There
is a growing  body of evidence  that it is  crucial to understanding
galaxy formation in  general.  Indeed the  light distribution  of most
large galaxies  is dominated by two components,   a bulge and  a disk. 
Even very early-type galaxies harbor a variety of luminosity profiles,
which are  interpreted  as due to  a  varying  contribution of a  disk
component  (Saglia et al.\ 1997;   de Jong et   al.  2004).  Along the
Hubble sequence, the variation in bulge magnitude  is twice the amount
of the disk (Simien \& de Vaucouleurs, 1986; de Jong, 1996), i.e.,
the properties of bulges are keys to inferring the nature of the Hubble
sequence.

Despite the prospect of yielding crucial information on galaxy
formation and galactic assembly history, bulges have  received
significantly less attention than elliptical galaxies.  This is a
direct consequence of the considerable challenge of avoiding disk
light contamination.  Figure \ref{LBLD} illustrates this point.
Galaxies, nearby and face-on, are from Jablonka, Arimoto \& Martin
(1996).  It appears clearly that within a fixed aperture, classically
of the order R$\sim$ 1-2 arcsec for integrated spectroscopy, it is
nearly impossible to get totally rid of the disk light.  Even more
importantly, one can get very different bulge-to-disk light ratios,
from one galaxy to the other, prejudicing our understanding of trends
with physical quantities.

Nevertheless, observational efforts intensify, improving our vision of
bulges' properties.  Although no definitive certainties have emerged
yet, new  lines of research are now underway.

\section{Metallicity distribution}

\begin{figure}
\begin{center}
\includegraphics[width=4.in,height=4.in]{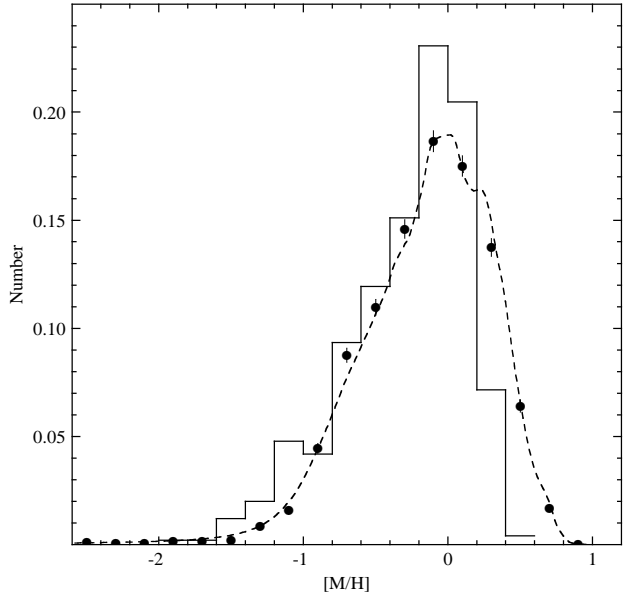}
\hfill
\caption{Comparison between the metallicity distribution of the  bulge of
M31  (filled points and   dotted line) and  the  one of the  Milky Way
(plain line), in regions  at comparable distances from the    galaxy
centers.}
\label{M31MDF}
\end{center}
\end{figure}

It will be a  long time before it is   possible to get  spectra   of
individual  stars in our  closest spiral neighbor,  M31.  Jablonka et
al.   (2000) using the MCS  deconvolution  technique counted $\sim$40
bright  RGB  stars per  arcsec$^2$ at   1 to 1.5~kpc  from  the galaxy
center.   Therefore,  attempts  to derive a   metallicity distribution
function  (MDF) in the bulge  of M31 have  to be based on high spatial
resolution images and analyzed with  isochrones.  At the moment, there
are two studies addressing this issue, one in the optical (Sarajedini
\& Jablonka, 2005), and the other in  the infrared (Olsen et al., 2006),
both based  on HST  data.   Sarajedini  \&  Jablonka analyze  a  field
located at  about 1.5~kpc from the nucleus  of M31 in  V and  I bands.
The MDF that they derive is presented in Figure
\ref{M31MDF} and compared   with the metallicity  distribution of  the
bulge  of the    Milky  Way from  Zoccali  et   al.  (2003).    Within
0.1-0.2~dex, the range of metallicity covered by the two galaxy bulges
is the same, and so are the peaks of the distributions.  Had bulges
straightforwardly reflected the differences between  the halos of  the
two galaxies, one would have expected a differential shift of at least
1~dex between  their MDFs (Ryan \&   Norris, 1991 ;  Durrell Harris \&
Pritchet, 2001).  On the contrary, it seems that the bulge of M31 does
not know  about the metal-richness  of its  halo.  Besides, the  M31
bulge MDF shows a  total absence of  metal-poor stars, just like does the
Milky Way  bulge.    This  is  a secure   result,   as the   lowest
metallicities identified  ($\sim -$1.5~dex)  are well away  from the
grey   zone   of  the  very  low    metallicities where isochrones are
degenerated in colors.

Contrary  to Sarajedini \& Jablonka, Olsen  et al (2006) leave the age
of the  stellar  population  as a  free  parameter  in their analysis.
Also,   instead of trying    to reproduce the complete color-magnitude
diagrams,  they  fit model  stellar  populations  to  the K luminosity
functions of their fields,  using a maximum  likelihood method.   They
find the stellar population mix of their 12 fields  to be dominated by
old (defined  as having ages  $\geq  $6 Gyr), nearly solar-metallicity
stars.  This old   population  seems to  dominate  the star  formation
history at  all radii, independent  of  the relative  contributions of
bulge   and disk stars.   In their  Figure~19, Olsen  et al.  show the
population integrated over all  their fields.  Neglecting the possibly
spurious   intermediate-age   metal-poor component,  possibly   due to
crowding,  they measure a  metallicity distribution function that is a
bit more sharply peaked than that of Sarajedini \& Jablonka, but still
in excellent qualitative agreement.

Assembling these  observational facts suggest that  the first stars in
bulges  formed from an already  pre-enriched  gas.  It remains unclear
whether this is resulting  from the halo  first stellar generations or
is due to the location of the observed fields.  Indeed, as we will see
later, bulges do exhibit radial gradients in metallicity and one might
not have yet probed  the outermost regions   of the M31 and Milky  Way
bulges.  In  any case, the bulk of  the bulge formation must have taken
place before  the mergers, whose  traces are witnessed today (Ferguson
et al.,  2002;  Ibata  et  al., 1994  ;  Yanny  et al.,  2003),  could
influence the bulge evolution.  Otherwise, the large difference between
the  M31  and  Milky Way halos  should    have been  reflected  in the
properties of their bulges.

\section{Scaling relations - central indices}

With the exception of the Milky  Way and M31,  in which we can resolve
individual   stars, studies of  bulges  have  to deal with  integrated
properties.

Spectroscopic studies of the central parts of bulges were pioneered by
Bica  (1988).  He gave the  first evidence for a relative independence
of  the bulge spectral  properties  with respect to the  morphological
type   of the parent  galaxies.    He  also  showed  that changes   in
age/metallicity  were linked to the  galaxy luminosity.  The following
years, a central  metallicity-luminosity  ($Z-L$) relation for  bulges
was more firmly established  and studies stressed its similarity  with
the  relation  derived for  ellipticals (Jablonka,  Martin \& Arimoto,
1996; Idiart,  de Freitas Pacheco   \& Costa, 1996).  This  similarity
appears both   in the  slope  of the  $Z-L$  relation  and in  similar
[$\alpha$/Fe] ratios.   Interestingly, both the above studies observed
face-on spirals  and  varied their  integration  apertures,  either by
adapting  their  spectroscopic   apertures    at  the time   of    the
observations, fixing a low and constant  bulge-to-disk light ratio for
all  galaxies,  or by inspecting   the light profiles  along  the slit
width when extracting the spectra.

Subsequent works sampled  inclined  galaxies and advocate  distinctions
between late-type and  early-type spiral bulges.  Prugniel, Maubon  \&
Simien (2001) bulges  are  located below the Mg$_2$-$\sigma$  relation
obtained  for  ellipticals.  Falc{\'o}n-Barroso,  Peletier \& Balcells
(2002) find a steeper  slope than for ellipticals  and S0  galaxies by
20\%.   Proctor  \& Sansom   (2002)  report  that  small bulges   (low
$\sigma$) depart  from    the relation between spectral    indices and
$\sigma$ drawn by large bulges:  While  large bulges populate the same
region as elliptical galaxies,  the smaller ones have relatively lower
spectral indices.   However,  Thomas  \& Davies  (2006),   reanalyzing
Proctor  and Sansom's sample, point out  that this apparent discrepancy
vanishes  when  the  same range   of central  velocity  dispersion  is
considered for  both types of  systems, i.e., when low $\sigma$ bulges
are compared to low $\sigma$ ellipticals.

Figure~4 in Falc{\'o}n-Barroso,   Peletier  \& Balcells (2002)   could
serve as a warning  : the dispersion  between the different studies is
rather large, likely due to the  various observational strategies.  In
particular, it is of  the  order of  the difference claimed  between
different types of bulges  and with elliptical galaxies. Nevertheless,
there  are true  points of  convergence among  the studies quoted here
which can be summarized as such: there is a range of properties of the
bulge stellar populations as sampled by  their inner regions.  They are
related to  the    bulge mass  or    maybe even   more to the    total
gravitational potential of the parent galaxy. Indeed, Prugniel, Maubon
\& Simien (2001)  and more recently Moorthy  \& Holtzman (2006) find a
tighter relation between Mg$_2$  and the galaxy rotation velocity than
with the  central bulge velocity  dispersion, for  example.  The bulge
central luminosity  weighted metallicities range   from $\sim -0.5$ to
$\sim +$  0.5 dex and the luminosity   weighted [$\alpha$/Fe] from the
solar value to $\sim$ 0.4 dex.  Ages are more subject to debate, but a
broad  consensus would certainly be reached  for a  range between very
old stellar population to a few giga years younger.

\section{Spatial distribution}

We would  dramatically limit our  understanding of bulges if one would
circumscribe the  analyses to their  central regions.  Substantial
progress   is enabled with  spatially   resolved spectroscopy, so that
radial  gradients    of   stellar   population   can    be   measured.
Investigations of such radial gradients using large surveys have until
recently only been addressed in  early-type galaxies i.e.,  elliptical
and  lenticular  galaxies, with only   very modest and rare excursions
into the case of later type galaxies (e.g., Sansom, Peace
\&  Dodd, 1994 ; Proctor,   Sansom \& Reid  , 2000  ;  Ganda et al.,
2006).

Recently,  Moorthy  \& Holtzman (2006)  published a  large study of 38
bulges, composed  for about half the sample  of nearly face-on spirals
and for the other half of highly inclined  ones.  Most of their bulges
show a steady  decrease in metallicity  sensitive indices  with radius
and positive increase in [$\alpha$/Fe],  with the exception that their
small  bulges have generally  weak or no gradients, sometimes positive
ones.   While age gradients    are generally absent  in their   sample
galaxies, some exhibit positive  ones, the majority in barred spirals. 
Very  interestingly, they  find   a correlation between  line strength
gradients in the bulge and in the disk.

\begin{figure}
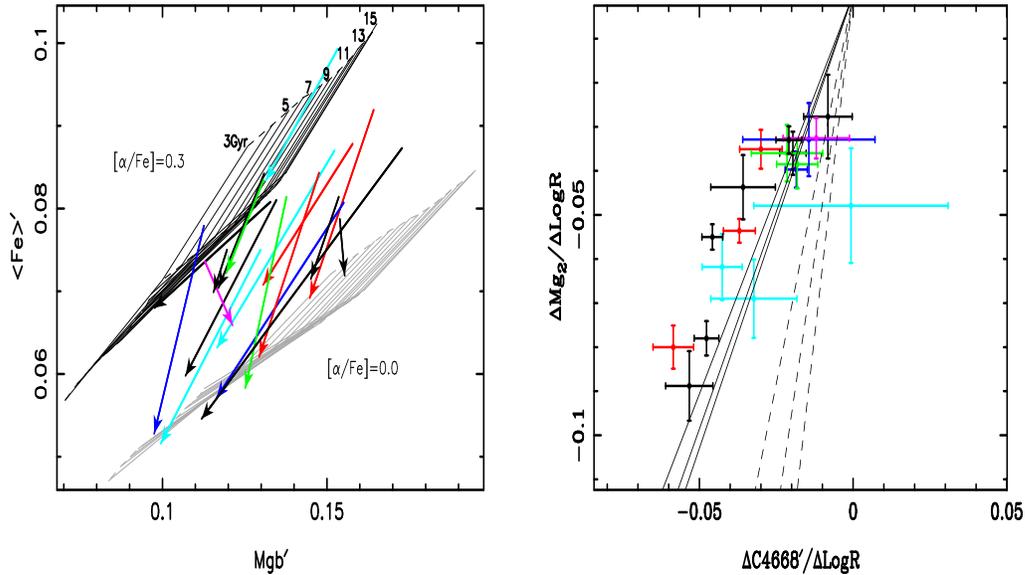

\includegraphics[width=3in,height=2.5in, angle=-90]{MgbFe.ps}
\hfill
\includegraphics[width=3in,height=2.5in, angle=-90]{Mg2C4668.ps}
\caption{Characteristics of bulges showing clear radial changes
  in  their spectral indices.  Colors are  coding the different Hubble
  types from S0 to Sbc. {\it Left panels}: The  arrows join the values
  of the indices $<$Fe$>$  and Mg$_b$ (expressed  in magnitude) at the
  bulge effective radius and in a central (r=2arcsec) aperture. Thomas
  et al.  (2003) grids of  single stellar population models are  shown
  at [$\alpha$/Fe]=0.0 and 0.3, for ages  between 3~Gyr and 15~Gyr and
  metallicities from  $-0.33$ to  $+0.35$.   {\it Right panel}:  Model
  lines of pure metallicity (plain lines) and  pure age (dotted lines)
  variations are derived.  They  are directly predicting the  relation
  between pairs of index gradients as observed in the bulge spectra.}
\label{Grad}
\end{figure}

Most of these qualitative results are confirmed by Jablonka, Gorgas \&
Goudfrooij  (2007)  who chose a different   strategy.  In order to get
totally rid of the disk population, they selected 32 genuine (or close
to)  edge-on spiral  galaxies with Hubble   types from S0 to Sc.  They
obtained spectra   along  the bulge  minor  axes,  out   to the  bulge
effective  radius  and often  much beyond.  Most   of their  bulges do
present radial stellar   population  gradients.  The outer   parts  of
bulges   do show weaker   metallic absorption   lines  than  the inner
regions.  The  distribution  of the gradient  amplitudes are generally
well peaked,  but they   also  display a  real intrinsic   dispersion,
implying the presence of a variety of star formation histories.

The  left  panel  of  Figure \ref{Grad}   illustrates the  decrease in
[$\alpha$/Fe],   from  the  bulge effective    radii to their  central
regions,  for galaxies  exhibiting  clear gradients.  The colors  code
different Hubble   types.   Alike the case    of the central  spectral
properties, the  morphology  of the parent galaxy   is  not a  driving
parameter  of  the  gradient  amplitudes.   The  right panel of Figure
\ref{Grad} shows the models of pure metallicity (plain lines) and pure
age variation (dotted lines)  from Thomas, Maraston \&  Bender (2003).
Bulges populate   the  region close   to the  pure  metallicity lines,
leaving some room, but  on a much  smaller magnitude for [$\alpha$/Fe]
and age  variations.   A quantitative analysis  indicates  that radial
gradients in luminosity-weighted mean metallicity are twice larger (in
logarithmic scale)   than the gradients in  age.   While [Fe/H] at the
bulge effective radii  is on average 0.4  dex lower than in the  bulge
central  regions, the age difference  is of the order  of 1.5 Gyr, the
inner  regions being younger.  The changes  in [$\alpha$/Fe] are small
(of the order of  0.1 dex) and rather   constant among bulges.   These
various points indicate that the outer regions  of bulges reveal their
earliest stages of star  formation.  Interestingly, the sensitivity of
the   gradients to the central  velocity  dispersion is very different
from what is reported for the bulge central indices.  Literally, there
is no correlation between the gradient amplitude and the bulge central
velocity dispersion.   Instead,   one  sees  that  bulges  with  large
velocity dispersions can  exhibit both strong or negligible gradients.
The probability to get  strong gradient diminishes at lower velocities
until it   gets  null.  The  same  had  been  observed  for elliptical
galaxies.   This gradual build-up of  the index vs.  $\sigma$ relation
can only be clearly  observed for indices  with large dynamical range,
such as Mg$_2$ or Mg$_1$.

\section{High redshift}

There are  still very  few studies   focusing  on bulges at  high
redshift, and only one is addressing  the comparison between bulges in
field and cluster environments.

Ellis, Abraham \& Dickinson (2001)  analyze a sample of early-type and
spiral galaxies from  the  northern and southern Hubble  Deep  Fields.
They compare the  central (inner 5\%) colors  of  spirals with clearly
visible bulges with   the  integrated colors of  ellipticals  in their
sample up to a redshift of $\sim$ 1.   They find that both ellipticals
and bulges show a dispersion in their colors  at a given redshift, but
that their  distributions   are  different  : a  smaller   fraction of
ellipticals is blue.   It seems that  there is an almost total absence
of bulges as red as those predicted by a passive evolution, while this
scenario provides on the contrary a good  description for the majority
of the early-type population.   The authors conclude that the  optical
luminosity weighted ages of bulges, to at least a redshift of 0.6, are
younger than those of   the   reddest ellipticals.  At   even   higher
redshifts  some bulges are found  to be as  red as ellipticals though,
suggesting that some kind  of rejuvenation is  at play at intermediate
redshift.

Koo et al. (2005a) present a sample of ellipticals and bulges from the
DEEP Groth Strip Survey with redshift  between $\sim$0.7 and $\sim$ 1.
This time, the images are decomposed into bulge and disk components by
fitting  a  de  Vaucouleurs light   profile   for the former  and   an
exponential one for  the latter. They  find that  red bulges (85\%  of
them) are nearly as red or redder  than the integrated color of either
local early-type  or  distant cluster galaxies.    The color-magnitude
relations  have similar shallow  slope and small scatter.  Blue bulges
are among  the least luminous ones, and  are  of similarly low surface
brightness as local bulges of similar size.  The authors consider that
they cannot be genuine proto-bulges and are instead mostly residing in
morphologically peculiar galaxies.  Interestingly, in most red objects,
they detect emission lines indicative of  a continuous star formation,
although at low level.

As stated by Koo   et al.  themselves,  the fact  that Ellis   et al.'s
sample encompasses faint bulges, while  their sample is restricted  to
luminous ones, together  with the very  different way  of deriving the
bulge colors  might  be at  the origin  of  the contradictory  results
between the two works.

Koo et al.   (2005b)  present the analysis  of luminous  bulge (MB $<$
19.5)  in a cluster and  in the  field at  redshift  $\sim$ 0.8.  They
demonstrate that the rest-frame  colors,  slope and dispersion of  the
color-magnitude of cluster and field bulges are nearly the same.  This
also means no larger  than in samples at lower   redshift. This is  in
sharp   contrast with some theoretical   expectation for an increasing
fraction of  recent star formation   with redshift and/or longer  time
scale of formation in the field as compared to clusters. However, here
again  the consequence  of the selection   of luminous bulges must  be
investigated.

\section{Conclusion}

This review definitely concentrated on the observational properties of
bulges.   These    were determined with     increasing accuracy  by an
ever-growing  number of bulge studies.   Here we  reviewed three major
areas    of   bulge   characterization: exploring     resolved stellar
populations, line indices from integrated spectra, and direct lookback
studies of bulge colors at redshifts out to  1.  Despite these diverse
approaches, however,  there is still a  large number  of viable models
for bulge formation, e.g.   gradual accretion of disk material through
the    action   of the     bars,  gravitational  collapse   (or nearly
equivalently, early and fast  mergers), the accretion of dwarf systems
in the center of disks, etc.  These models  are not mutually exclusive
and in reality bulge growth may combine some of these modes.  The lack
of agreement in  the conclusions between  some analyses mentioned here
serves to  guide us towards  questions that remain open.  Among those:
at which point  in their history do  the disks influence their central
bulges ?  If they do, then in which  proportion?  Could this influence
become  dominant in small  systems ?   Do  we have/use the appropriate
comparison    samples of   elliptical  galaxies  ?   Which gravitation
potential governs most closely the formation of  bulges: theirs or the
total galaxy one ?

\end{document}